\documentclass[twocolumn,pra,superscriptaddress,showpacs,aps]{revtex4-1}
\usepackage{color}
\usepackage{amsmath}
\usepackage{amssymb}
\usepackage{txfonts}
\usepackage{graphicx}
\usepackage[unicode]{hyperref}
\hypersetup{pdftitle={Breather},
 pdfauthor={S-W. Su, S-C. Gou, I-K. Liu, A. S. Bradley, S. S. Shamailov, O. Fialko, J. Brand},
 pdfsubject={Sine Gordon Breather},
    colorlinks=true,
    linkcolor=blue,
    citecolor=blue,
    filecolor=black,
    urlcolor=blue
}

\usepackage{epstopdf}

\DeclareMathOperator{\erf}{erf}

\begin{document}

\title{Oscillons in coupled Bose-Einstein condensates}

\author{Shih-Wei Su}

\affiliation{Department of Physics and Graduate Institute of Photonics, National
Changhua University of Education, Changhua 50058 Taiwan}

\author{Shih-Chuan Gou}

\affiliation{Department of Physics and Graduate Institute of Photonics, National
Changhua University of Education, Changhua 50058 Taiwan}

\author{I-Kang Liu}

\affiliation{Department of Physics and Graduate Institute of Photonics, National
Changhua University of Education, Changhua 50058 Taiwan}

\author{Ashton S. Bradley}

\affiliation{Dodd-Walls Centre for Photonics and Quantum Technology, Department of Physics, University
of Otago, Dunedin, New Zealand}

\author{Oleksandr Fialko}

\affiliation{Centre for Theoretical Chemistry and Physics, Institute for Natural and Mathematical Sciences, Massey University, Auckland, New Zealand}

\author{Joachim Brand}

\affiliation{Dodd-Walls Centre for Photonics and Quantum Technology and Centre for Theoretical Chemistry and Physics, New Zealand Institute
for Advanced Study, Massey University, Auckland, New Zealand}

\begin{abstract}
Long-lived, spatially localized, and temporally oscillating nonlinear
excitations are predicted by numerical simulation of coupled Gross-Pitaevskii
equations. These oscillons closely resemble the time-periodic breather
solutions of the sine-Gordon equation but decay slowly by radiating
Bogoliubov phonons. Their time-dependent profile is closely matched
with solutions of the sine-Gordon equation, which emerges as
an effective field theory for the relative phase of two linearly coupled Bose fields
in the weak-coupling limit. For strong coupling the long-lived oscillons
persist and involve both relative and total phase fields. The oscillons
decay via Bogoliubov phonon radiation that is increasingly suppressed
for decreasing oscillon amplitude. Possibilities for creating
oscillons are addressed in atomic gas experiments by collision of oppositely charged
Bose-Josephson vortices and direct phase imprinting.  

\end{abstract}


\maketitle

\section{INTRODUCTION}

Oscillons are localized and oscillating concentrations of energy in 
a nonlinear field that decay very slowly over many of their oscillation
periods \cite{Bogolyubskii1976,Gleiser1994,Vachaspati2006}. They
are thought to be relevant in the dynamics of cosmological phase transitions~\cite{Copeland1995} 
and inflation scenarios~\cite{Amin2012}, where
they provide localized stores of energy. In contrast to cosmic strings~\cite{Hindmarsh1995} 
or other topological defects that can only be
destroyed by annihilation, isolated oscillons can decay by radiation
over long time scales. Closely related to oscillons are breathers,
which are time-periodic localized solutions of a nonlinear field equation~\cite{Vachaspati2006}. 
Breathers only exist when resonance between
the breather frequency and extended linear waves is avoided. This
is known to happen only in integrable nonlinear wave equations like
the sine-Gordon (SG) equation~\cite{Ablowitz1973} and the nonlinear Schr\"{o}dinger
equation~\cite{Akhmediev1987}, or in nonlinear lattices~\cite{Flach1998}
due to the occurrence of band gaps in the linear wave spectrum. Oscillons
are thus the more generic localized and oscillating nonlinear excitations.

In most nonlinear field equations, oscillon solutions are not known in closed form and the investigation of oscillon properties relies on numerical simulations. However, due to the integrability
of SG equation in $1+1$ dimensions~\cite{Bour,Frenkel}, the closed form solutions of SG breathers as well as solitons can be derived~\cite{Physofsoliton}. 
The fundamental solitons of the SG equation are localized topological excitations with a topological charge of $\pm 1$  known as kinks and anti-kinks, respectively. Sine-Gordon breathers are localized in space and 
periodic in time and can be understood as the bound states of kink and anti-kink. 
Since their total topological charge is zero, breathers are not protected by their topological properties but rather by the integrable nature
of the SG equation. Slight modifications of the equation that break
the integrability typically destroys the breather solutions. However,
numerical methods can still find long-lived spatially localized
and oscillating concentrations of energy, namely oscillons~\cite{Peyrard1983,Vachaspati2006,Abdullaev2013}.

Linearly coupled BECs have been proposed as a model system to simulate
the sine-Gordon equation in a variety of contexts~\cite{Gritsev2007,Neuenhahn2012,Opanchuk2013}. 
This has aroused considerable interest in investigating the excitations of linearly coupled Bose gases,  in particular Josephson vortices, which are closely related to SG kinks~\cite{Kaurov2005,Kaurov2006,Brand2009,Qadir2012,shamailov14}.
In a recent paper several of us have simulated the spontaneous formation and decay of Josephson vortices within the Kibble-Zurek scenario of a rapid quench through the BEC phase transition in quasi-one-dimensional coupled Bose gases~\cite{Su2013}. The finding that post-quench dynamics and in particular the collision and annihilation of Josephson vortices contribute decisively to the breakdown of  Kibble-Zurek scaling laws has motivated the current study. As shown in Sec.~\ref{S:collison}, the collision of slow and oppositely charged Josephson vortices generically produces oscillons.
  
It has further been suggested that breather-like excitations (oscillons) with a finite lifetime can be spontaneously formed in coupled BECs via a dynamical instability triggered
by parametric amplification of quantum fluctuations~\cite{Neuenhahn2012}.
This emergence of localized excitations from amplified quantum fluctuations
bears close analogy with the formation of oscillons after inflation of the early universe predicted within relativistic cosmological
models~\cite{Amin2012}. The coupled BECs may thus provide an experimentally accessible model to simulate the evolution of the early universe~\cite{Opanchuk2013}. 

While recent works on simulating the SG model using coupled BECs have focussed entirely on the relative phase dynamics~\cite{Gritsev2007,Neuenhahn2012,Opanchuk2013}, this picture neglects
the coupling to the symmetric degrees
of freedom in the system, such as the total phase and densities of
the condensates. Realistically, such couplings can serve as channels
of decoherence that can significantly alter the relative-phase dynamics 
and then lead to the instability of non-topological excitations, such as breathers.
A similar circumstance arises, for example, in annular Josephson junctions,
where an effective dissipation is introduced by carefully eliminating
electronic degrees of freedom~\cite{Ambegaokar,Eckern}. By the same
token, it has been shown that the breathers are not robust against an ambient perturbation and hence decay
by continuously radiating phonons~\cite{Harvey,Boyd,Denzler} thus 
forming oscillons. This motivates us to carry out a detailed study of the full dynamics of
breather-like excitations in BECs. Specifically, we investigate the
time evolution of an initial SG-breather state by numerically solving
the Gross-Pitaevskii equation over broad ranges of coupling energy and imprinted SG breather
frequency. We assess the stability of the long-lived oscillon, and
present a simple and feasible experimental scheme to create an oscillon
by phase-imprinting method in a double-ring BEC system. 

The organization of this paper is as follows. In Sec. II, we give
a brief account of the system consisting of two linearly coupled BECs.
We outline the emergent SG sector in the weak-coupling limit, and its
kink solutions. In Sec. III, we numerically demonstrate that the oscillon
forms by imprinting the SG breather onto the initial state of the
BECs, and the stability of the ocillon is studied for broad ranges of coupling
energy and imprinted SG breather frequency. In Sec. IV, we demonstrate the
creation of an oscillon from the collision of two Josephson vortices,
and outline experimental constraints for a realistic phase imprinting
protocol in a double-ring geometry. We conclude in Sec. V. 

\section{MODEL}

We consider a system consisting of two linearly coupled BECs, which
are tightly confined in the transverse directions but loosely confined
in the longitudinal direction. The grand-canonical Gross-Pitaevskii
(GP) energy functional for this quasi-one-dimensional system is given
by 
\begin{eqnarray}
E & = & \sum_{j=1,2}\int dz\left[\frac{\hbar^{2}}{2m}\left\vert \partial_{z}\psi_{j}\left(z\right)\right\vert ^{2}+\frac{g}{2}\left\vert \psi_{j}\left(z\right)\right\vert ^{4}-\mu\left\vert\psi_{j}\right\vert^{2}\right]\notag\\
 &  & -\tilde{\nu}\int dz\left[\psi_{1}^{\ast}\left(z\right)\psi_{2}\left(z\right)+\psi_{2}^{\ast}\left(z\right)\psi_{1}\left(z\right)\right],\label{Grand-EGP}
\end{eqnarray}
where $\psi_{j}$ is the order parameter of the \textit{\textcolor{black}{j}}-th
condensate, $\mu$ the chemical potential, $\tilde{\nu}$ the coupling
strength characterizing the tunnelling energy between the two systems,
and $g$ is the effective nonlinear interaction strength. The equation
of motion for each $\psi_{j}$ can be derived via the Hartree variational
principle, namely, $i\hbar\partial_{t}\psi_{j}=\delta E/\delta\psi_{j}^{\ast}$.
For computational simplicity, we let the length and time to be scaled
in units of $l_{c}=\hbar/\sqrt{m\mu}$ and $t_{0}=\hbar/\mu$ respectively,
so that the coupled GP equations are expressed in the dimensionless
form 
\begin{align}
i\partial_{t}\psi_{1} & =\left(-\partial_{z}^{2}/2+\left\vert \psi_{1}\right\vert ^{2}-1\right)\psi_{1}-\nu\psi_{2},\nonumber \\
i\partial_{t}\psi_{2} & =\left(-\partial_{z}^{2}/2+\left\vert \psi_{2}\right\vert ^{2}-1\right)\psi_{2}-\nu\psi_{1},\label{eq:GPE}
\end{align}
where $\nu=\tilde{\nu}/\mu$ is the dimensionless coupling energy.

The ground state of the coupled BECs can be determined by minimizing
the energy functional, Eq.~(\ref{Grand-EGP}), where we seek the minimum
of $V(\psi_{1},\psi_{2})\equiv\sum_{j=1,2}|\psi_{j}|^{2}\left[\frac{1}{2}|\psi_{j}|^{2}-1\right]-\nu\left[\psi_{1}^{*}\psi_{2}+\psi_{2}^{*}\psi_{1}\right]$.
The symmetry $V(\psi_{1},\psi_{2})=V(\psi_{2},\psi_{1})$ imposes
a common amplitude for the ground state fields. Taking $\psi_{1}^{0}=\sqrt{\rho_{0}}e^{i\phi_{1}}$,
$\psi_{2}^{0}=\sqrt{\rho_{0}}e^{i\phi_{2}}$, and $\Delta=\phi_{1}-\phi_{2}$,
the minimum of $V=\rho_{0}^{2}-2\mu\rho_{0}-2\nu\rho_{0}\cos{\Delta}$
occurs at $\Delta=0$ and $\rho_{0}=(1+\nu)$.

\subsection{Collective Excitations}

To find the low-lying excitations above the ground state, we can replace
$\psi_{j}$ with $\psi_{j}^{0}+\alpha_{q}^{j}e^{iqz-i\omega t}-\beta_{q}^{j*}e^{-iqz+i\omega t}$
in Eq.~(\ref{eq:GPE}) and retain the expressions up to the linear
order in $\alpha_{q}$ and $\beta_{q}$ , which then leads to the
Bogoliubov-de Genes (BdG) equation 
\begin{equation}
\left[\begin{array}{cccc}
H_{0}-\omega & -\left(1+\nu\right) & -\nu & 0\\
1+\nu & -H_{0}-\omega & 0 & \nu\\
-\nu & 0 & H_{0}-\omega & -\left(1+\nu\right)\\
0 & \nu & 1+\nu & -H_{0}-\omega
\end{array}\right]\left[\begin{array}{c}
\alpha_{q}^{1}\\
\beta_{q}^{1}\\
\alpha_{q}^{2}\\
\beta_{q}^{2}
\end{array}\right]=0,\label{BdG}
\end{equation}
where $H_{0}=q^{2}/2+(1+2\nu)$. The low-lying excitation spectrum
is determined by solving the eigenvalue problem, Eq.~(\ref{BdG}).
As a result, we obtain two distinct dispersion relations for the excited
modes: 
\begin{eqnarray}
\omega_{1} & = & \sqrt{\frac{q^{2}}{2}\left(\frac{q^{2}}{2}+2\nu+2\right)},\label{gapless eigenvalue}
\end{eqnarray}
with the eigenvector $\sim$ $(u,v,u,v)^{T}$, and 
\begin{eqnarray}
\omega_{2} & = & \sqrt{\left(\frac{q^{2}}{2}+2\nu\right)\left(\frac{q^{2}}{2}+4\nu+2\right)},\label{gapped eigenvalue}
\end{eqnarray}
with the eigenvector $\sim$ $(u,-v,-u,v)^{T}$. Equation~(\ref{gapless eigenvalue})
represents a gapless mode, corresponding to the in-phase excitation
of the two components of condensates, which manifests itself as the
Bogoliubov sound wave propagating at a speed $v_{Bog}=\sqrt{1+\nu}$.
On the other hand, Eq.~(\ref{gapped eigenvalue}) indicates a gapped
mode, which corresponds to out-of-phase excitation. It should
be noted that these two modes are decoupled in the linear approximation.
Moreover, the gapped excitation possesses a relativistic energy dispersion
$\omega_{2}^{2}=q^{2}c_{2}^{2}+m_{2}^{2}c_{2}^{4}$ with sound
speed, $c_{2}^{2}=1+3\nu$ and the rest energy, $m_{2}^{2}c_{2}^{4}=4\nu(1+2\nu)$.
The gapless branch accounts for the relative phase dynamics which can be described
by the sine-Gordon equation~\cite{Opanchuk2013}. 
\begin{figure*}
\includegraphics[width=1.7\columnwidth]{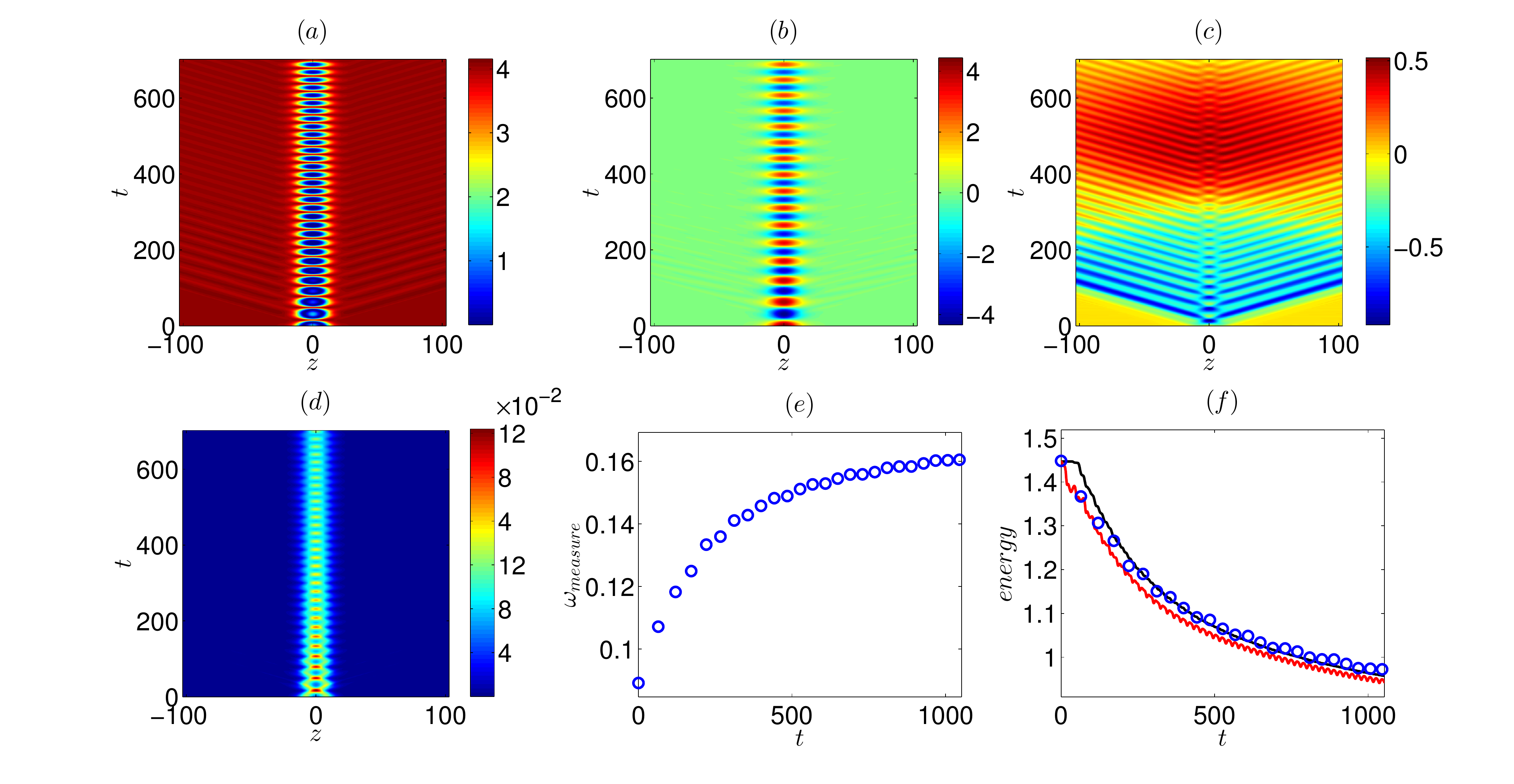} \protect\caption{(Color online) Time evolution of the oscillon with imprinted SG-breather parameters, $u=0.5$ and $\nu=0.01$ [Eq.~(\ref{breather})]:
(a) The density profile of the two superposed atomic fields, $|\psi_{1}+\psi_{2}|^{2}$.
(b) The spatial distribution of the relative phase, $\phi_{1}-\phi_{2}$.
(c) The spatial distribution of the total phase, $\phi_{1}+\phi_{2}$.
(d) The density profile of the grand canonical GP energy $E$. (e) The
measured frequency of the oscillon. (f) The computed localized energy
of the oscillon: blue circle indicates the energy $E_B$ obtained according
to Eq.~(\protect\ref{Breather-energy}); black line indicates the
energy $E$ obtained by integrating Eq.\ (\protect\ref{Grand-EGP}); red
line indicates the energy $H_{\mathrm{\textrm{SG}}}$ obtained by integrating Eq.~(\protect\ref{SG-GP-subset}).}
\label{nu1e-2-u-5e-1} 
\end{figure*}

\subsection{Sine-Gordon Regime}

Before studying the oscillon excitation in linearly coupled BECs,
we investigate the low-energy dynamics of the symmetric degrees of
freedom. In general, the order parameters of the coupled BECs can
be represented by $\psi_{1}=\sqrt{\rho}e^{i(\phi_{s}+\phi_{a})/2}\sin\theta$
and $\psi_{2}=\sqrt{\rho}e^{i(\phi_{s}-\phi_{a})/2}\cos\theta$, where
$\phi_{s}$ and $\phi_{a}$ are respectively the total and relative
phase, $\rho$ the total density, and $\theta$ is the density mixing
angle. Substituting $\psi_{1}$ and $\psi_{2}$ into Eq.~(\ref{Grand-EGP})
with the assumption that the coupling between the two BECs is sufficiently
weak, and following the arguments used in Ref.~\cite{Opanchuk2013}, we obtain
the relevant Hamiltonian density accounting for the sine-Gordon dynamics
\begin{eqnarray}
H_{\mathrm{\textrm{SG}}} & = & \frac{\left(1+\nu\right)}{4}\left(\partial_{z}\phi_{a}\right)^{2}+\left(1+\nu\right)^{2}\left(1+\cos4\theta\right)\notag\\
 &  & -2\nu\left(1+\nu\right)\left(\cos\phi_{a}-1\right)\sin2\theta,\label{SG-GP-subset}
\end{eqnarray}
where the total density $\rho$ is replaced by the sum of the ground
state density of the two components, namely, $2\rho_{0}=2\left(1+\nu\right)$.
Using the variational principle, the Euler-Lagrange equation of the
relative phase is derived~\cite{Opanchuk2013}, 
\begin{equation}
\partial_{t}^{2}\phi_{a}-(1+\nu)\partial_{z}^{2}\phi_{a}+4\nu\left(1+\nu\right)\sin\phi_{a}=0,\label{sine-Gordon}
\end{equation}
which is equivalent to the SG equation.
It is well known that the SG equation possesses the topologically protected kink and antikink solutions~\cite{Physofsoliton}, i.e., 
\begin{eqnarray}
\phi_{a}^{S} & = & 4\tan^{-1}\left[\exp\left(\pm\sqrt{4\nu}\frac{z-z_{0}}{\sqrt{1-2v^{2}/\left(1+\nu\right)}}\right)\right],\label{soliton}
\end{eqnarray}
where the $+$$(-)$ sign corresponds to the kink (antikink) and $v$ and $z_{0}$
are the velocity and the initial location of the (anti)kink, respectively.
Remarkably, the kink and antikink in Eq.~(\ref{soliton}) may form
a bound kink-antikink pair. Such a solution does exist for the SG
model, which is spatially localized but temporally oscillating, and
is thus called breather, 
\begin{eqnarray}
\phi_{a}^{B} & = & 4\tan^{-1}\left[\frac{\sqrt{1+\nu}\sin\sqrt{\frac{4\nu}{1+u^{2}/\left(1+\nu\right)}}ut}{u\cosh\sqrt{\frac{4\nu}{1+u^{2}/\left(1+\nu\right)}}z}\right].\label{breather}
\end{eqnarray}
Here $u$ is a positive constant that parametrises the amplitude
and frequency of the breather by 
\begin{equation}
\phi_{a,max}^{B}=4\tan^{-1}\left(\frac{\sqrt{1+\nu}}{u}\right), \label{breathfreq}
\end{equation}
and
\begin{equation}
 \omega_{B}=u\sqrt{\frac{4\nu}{1+\frac{u^{2}}{1+\nu}}}, 
\end{equation}
respectively. In
contrast to kink and antiknik, the breather is not protected by topology
but is protected by the integrability of the SG equation. The breather
becomes unstable once the integrability is broken. From Eq.~(\ref{breather}), we can define the full width at half maximum
(FWHM) of the breather, $z_{w}=\sqrt{(1+u^{2}/(1+\nu))/4\nu} \mathrm{sech}^{-1}(1/2)$.
Note that the amplitude (frequency) decreases (increases) with increasing
$u$. With Eq.~(\ref{breather}), it is straightforward to show that
the energy of a breather is given by 
\begin{eqnarray}
E_{B} & =16(1+\nu)\sqrt{\nu\left(1-\tilde{\omega}^{2}\right)},\label{Breather-energy}
\end{eqnarray}
with $\tilde{\omega}=\omega_{B}/\sqrt{4\nu(1+\nu)}$~\cite{Samuelsen}.
In addition to the kink and breather, the SG equation also possesses
extended phonon excitations \cite{Physofsoliton}.

The Hamiltonian, Eq.~(\ref{SG-GP-subset}), represents an emergent
SG sector in the weak coupling limit of the total GP Hamiltonian, where the latter supports excitations described by both
asymmetric and symmetric degrees of freedom. Excitations
in the SG sector such as the kink, breather, and phonon, depend only
on the asymmetric degrees of freedom. On the other hand, off the SG
sector, there are other types of excitations depending on the symmetric
degrees of freedom, such as the Bogoliubov phonon (gapless excitation),
and the GP dark (grey) soliton which has a $2\pi$$\left(<2\pi\right)$
phase jump in the total phase. Under the framework of the full GP
formalism, theses excitations of different degrees of freedoms would
couple to each other and these couplings would play an important role
in the dynamics of the condensates. Therefore, it is expected that
the SG breather is not stable in the presence of these couplings and
would decay by slowly radiating energy, thus forming a so-called oscillon.
In the following section, we investigate dynamics of the oscillon
excitations in the linearly coupled BECs. 

\section{Dynamical stability}

In order to create an ocillon, we begin with a phase imprinted SG breather,
with relative phase given by Eq.~(\ref{breather}), initially imprinted
into a system of two linearly-coupled BECs in a double-ring geometry.
The couplings between different degrees of freedom lead to the instability
of the initially imprinted breather and result in energy radiation
in terms of sound waves. We expect that the propagation of the emitted
sound waves may introduce extra effects to the dynamics of the oscillon,
and our aim is to investigate how the oscillon evolves under the influence
of the evolving background condensates.

A technical issue arrises, namely that, since we impose periodic boundary
condition corresponding to a double-ring
geometry for the numerical computations, whenever the outgoing sound waves reach one of the boundaries
they will re-enter the system from the boundary at the opposite side.
These re-entering waves will interfere with those outgoing ones and
also interact with the oscillon, complicating the dynamics in an uncontrollable
way. To suppress the reentry of the emitted sound waves, we impose
the absorption boundary layers by including a position-dependent damping
coefficient $\sigma\left(z\right)$ in Eq.~(\ref{eq:GPE}), which turns
Eq.~(\ref{eq:GPE}) into a damped GP equation~\cite{Penckwitt,Tsubota,Billam}
\begin{eqnarray}
i\partial_{t}\psi_{1} & = & \left(1-i\sigma\right)\left[\left(-\partial_{z}^{2}/2+\left\vert \psi_{1}\right\vert ^{2}-1\right)\psi_{1}-\nu\psi_{2}\right],\label{eq:DGPE-1} \\
i\partial_{t}\psi_{2} & = & \left(1-i\sigma\right)\left[\left(-\partial_{z}^{2}/2+\left\vert \psi_{2}\right\vert ^{2}-1\right)\psi_{2}-\nu\psi_{1}\right].\label{eq:DGPE-2}
\end{eqnarray}
The damping coefficient $\sigma$ varies with $z$ according to the
prescription, $\sigma(z)=1+[\erf(s(z-z_{0}))-\erf(s(z+z_{0}))]/2$,
where $z_{0}$ and $1/s$ are the center and the width of the error
function, respectively. In the simulations, $z_{0}$ is chosen to
locate far from the origin and $s$ is set to be sufficiently small,
such that $\sigma$ is increasing smoothly over the boundary layer
and the reflected sound waves are greatly reduced. From a physical
point of view, the absorber means that we are effectively considering
a very large ring system, where the feedback from acoustic emissions
lags the oscillon dynamics by a significant timescale. 

In what follows, we present the main results obtained by numerically
integrating Eqs.~(\ref{eq:DGPE-1}) and~(\ref{eq:DGPE-2}) over a variety of initial conditions,
where the space integration by highly accurate Fourier pseudospectral
method, and the time integration is by the adaptive Runge-Kutta method
of orders 4 and 5 (RK45). 

\subsection{Dynamics of the Oscillon}

In order to study the dynamics and stability of oscillons in the coupled BECs, we imprint the phase
profile of the SG breather, $\phi_{a}^{B}$, which is defined in Eq.~(\ref{breather}), onto the homogeneous background density $n_{1}^{0}=n_{2}^{0}=1+\nu$,
\begin{eqnarray}
\psi_{1}\left(t=0\right) & =\psi_{2}^{\ast}\left(t=0\right)=\sqrt{1+\nu}e^{i\phi_{a}^{B}/2},\label{initial-state}
\end{eqnarray}
and evolve this initial state in accordance with Eqs.~(\ref{eq:DGPE-1}), (\ref{eq:DGPE-2}).
\begin{figure}[!t]
\includegraphics[width=1\columnwidth]{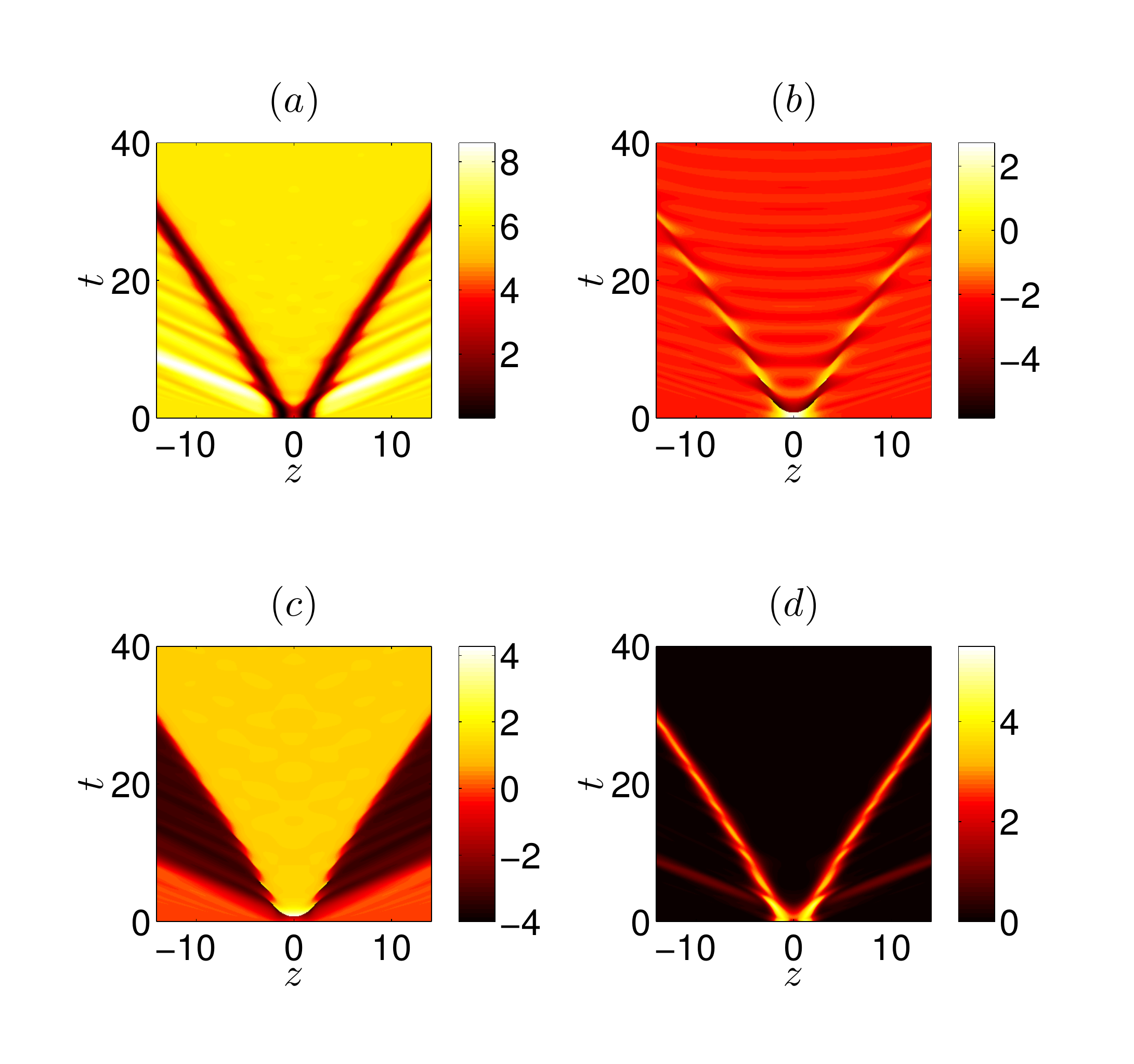} \protect\caption{(Color online) Formation of GP solitons with the imprinted SG-breather parameters, $u=0.5$ and $\nu=0.5$ [Eq.~(\ref{breather})]: (a) The density
profile of the two superposed atomic fields, $|\psi_{1}+\psi_{2}|^{2}$.
(b) The spatial distribution of the relative phase, $\phi_{1}-\phi_{2}$.
(c) The spatial distribution of the total phase, $\phi_{1}+\phi_{2}$.
(d) The density profile of the grand canonical GP energy $E$. Note that
the imprinted SG breather is unstable which instantly decays into two GP solitons. }
\label{nu-5e-1-u-5e-1} 
\end{figure}
\begin{figure*}[htbp]
\includegraphics[width=2\columnwidth]{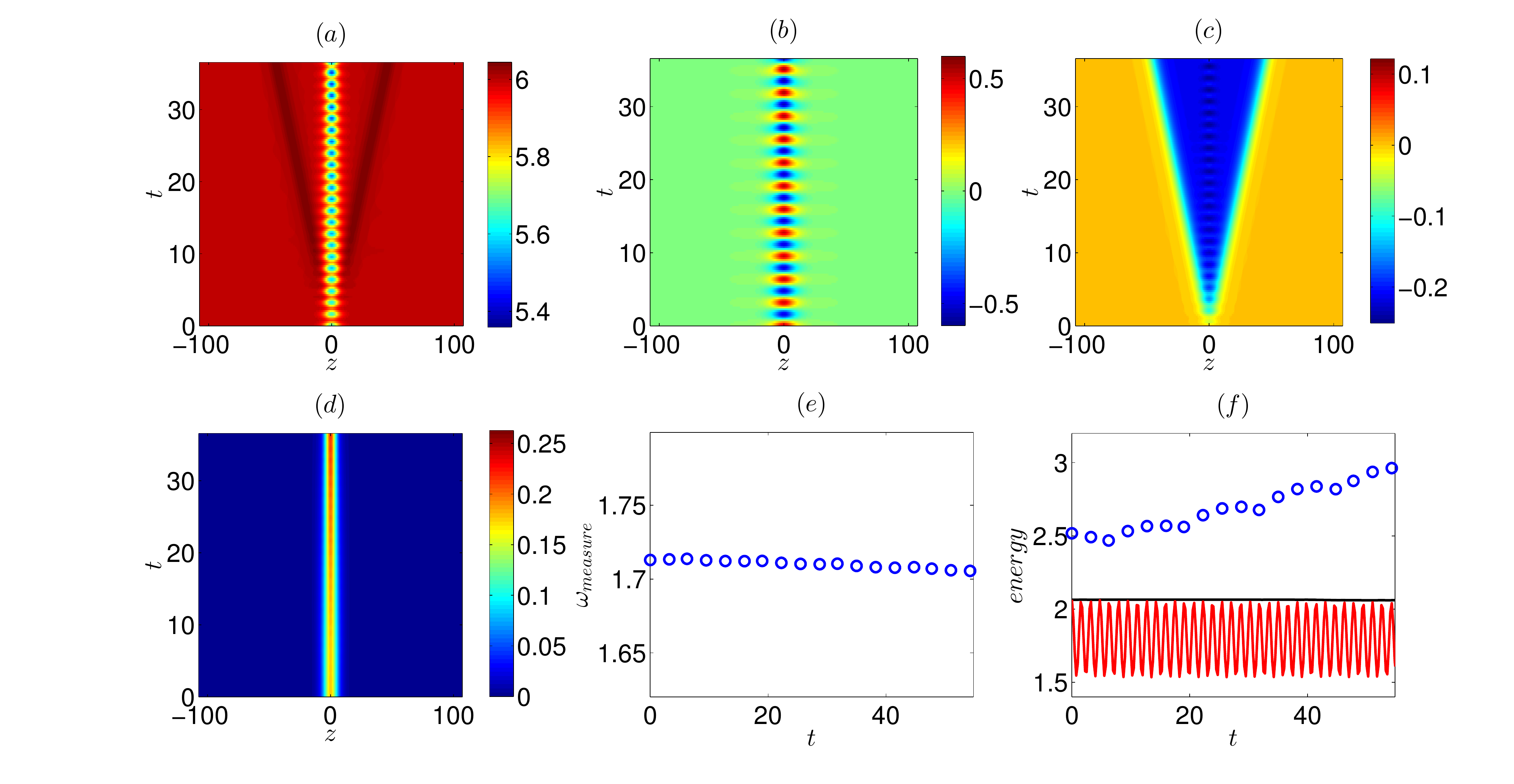} \protect\caption{(Color online) Time evolution of the oscillon with imprinted SG-breather parameters, $u=10$ and $\nu=0.5$ [Eq.~(\ref{breather})]: (a) The
density profile of the two superposed atomic fields, $|\psi_{1}+\psi_{2}|^{2}$.
(b) The spatial distribution of the relative phase, $\phi_{1}-\phi_{2}$.
(c) The spatial distribution of the total phase, $\phi_{1}+\phi_{2}$.
(d) The density profile of the grand canonical GP energy $E$. (e) The
measured frequency of the oscillon. (f)
The computed localized energy
of the oscillon: blue circle indicates the energy $E_B$ obtained according
to Eq.~(\protect\ref{Breather-energy}); black line indicates the
energy $E$ obtained by integrating Eq.\ (\protect\ref{Grand-EGP}); red
line indicates the energy $H_{\mathrm{\textrm{SG}}}$ obtained by integrating Eq.~(\protect\ref{SG-GP-subset}).
The deviation of the red and black lines in panel (f) indicates that the
strongly-coupled condensates cannot be described by the SG equation
for the relative phase. }
\label{nu5e-1-u10} 
\end{figure*}

\subsubsection{Weak Coupling}

We first consider the case of a weak-coupling energy $\nu=0.01$,
and $u=0.5$ for the initially imprinted SG breather profile. The created
oscillon can be easily identified in the pattern of superposition,
$|\psi_{1}+\psi_{2}|^{2}$ and in the time-varying spatial distribution
of the relative phase, $\phi_{1}-\phi_{2}$, as shown in Fig.~\ref{nu1e-2-u-5e-1}(b).
Continuous emission of Bogoliubov sound waves from the oscillon is
revealed by the appearance of the wavefronts in Fig.~\ref{nu1e-2-u-5e-1}(a)
and (c), that corresponds to the up-chirping of the oscillon
frequency as shown in Fig.~\ref{nu1e-2-u-5e-1}(e). To characterize
the evolution of the oscillon, we measure the frequency (period) of
the relative phase oscillation shown in Fig.~\ref{nu1e-2-u-5e-1}(e).
To verify the degree of deviation of the oscillon from the SG breather,
we compare the energy of the oscillon with those calculated by integrating the GP energy functional
(\ref{Grand-EGP}) and SG Hamiltonian (\ref{SG-GP-subset}) within the region
of localization of the oscillon, $-7z_{w}<z<7z_{w}$. As shown in
Fig.~\ref{nu1e-2-u-5e-1}(f), the energies $E_B$ of Eq.\ (\ref{Breather-energy}) obtained from the measured oscillon frequency
(blue circle), the GP energy $E$ of Eq.~(\ref{Grand-EGP}) (black line), and the SG Hamiltonian $H_{\mathrm{\textrm{SG}}}$ of  Eq.~(\ref{SG-GP-subset})
(red line) agree closely with each other, implying that the resulting
oscillon is almost identical to the SG breather. Furthermore, the red
line shows small-amplitude oscillation which is owing to the coupling
of the asymmetric degrees of freedom to the symmetric ones and this
suggests that the observed localized energy oscillation is in the
context of oscillon. We also perform the simulations for much smaller
couplings and the oscillation of SG energy and the emission of the
Bogoliubov sound are suppressed which suggests that for sufficiently
weak coupling ($\nu\ll1$) the coupled BECs support long-lived
oscillon-type excitations. Studies of both the $\phi^{4}$ and the perturbed SG models have predicted the decay of the oscillon excitation
via phonon emission~\cite{Harvey,Boyd,Denzler}. However
in the weakly coupled BEC system we observe a slightly different decaying
behaviour where the oscillon-type excitations in the asymmetric degrees
of freedom lose energy to the symmetric degrees of freedom by emitting
Bogoliubov phonons. Unlike the $\phi^{4}$ and SG model, the decay
which we observed could only occur in the coupled BECs since the coupled
BECs support different degrees of freedom. 

\subsubsection{Strong Coupling} \label{S:strong-coupling}

In the strong-coupling regime, the relative phase dynamics is beyond
the SG description so the dynamical properties of the oscillon are
expected to be different from those of the SG breather. To study the
oscillon dynamics in the strong coupling regime, we follow the same
simulation procedure in the weakly coupled BECs by imprinting the
SG breather onto the ground state. We first consider a stronger coupling
energy of $\nu=0.5$ and an initially imprinted SG breather of $u=0.5$.
The evolution is shown in Fig.~\ref{nu-5e-1-u-5e-1}, where the imprinted
SG breather is unstable and decays into two GP solitons instantly.
We see that the two GP grey solitons, which carry the majority of
the excitation energy, move towards the boundaries of the coupled
condensates. However, we find that the oscillon could remain long-lived
by increasing $u$ of the initially imprinted SG breather. Now let
us consider the same coupling energy but with an otherwise different
initial imprinted SG breather of $u=10$. The time evolution of this
initial state is shown in Fig.~\ref{nu5e-1-u10}. In Fig.~\ref{nu5e-1-u10}(a),
the imprinted SG breather emits sound waves initially to reduce its
energy and then form the long-lived oscillon excitation. In contrast
to the ocillon in the SG regime, the oscillon in the strongly coupled
BECs exhibits different dynamical behaviour - the oscillon only emits
sound waves at the beginning and subsequently the total phase
manifests localized breathing oscillation as shown in Figs.~\ref{nu5e-1-u10}(a)
and (b). Similar to the weak coupling case, we also compare the energies
of the oscillon obtained by Eqs.~(\ref{Grand-EGP}), (\ref{SG-GP-subset}),
and (\ref{Breather-energy}), and find that all these results are
noticeably different. We see that the GP energy density {[}Fig.~\ref{nu5e-1-u10}(d){]} shows a different behaviour from the weak coupling
case {[}Fig.~\ref{nu1e-2-u-5e-1}(d){]}. The latter exhibits a
temporally oscillatory behaviour but the former does not. This oscillation
implies the energy transfer between the asymmetric degrees of freedom
and the symmetric degrees of freedom which also accounts for the emergence
of the oscillon in both the total and relative phases. Furthermore, the
deviation of GP energy from the breather energy obtained by inserting
the measured frequency into Eq.~(\ref{Breather-energy}) {[}Fig.~\ref{nu5e-1-u10}(e){]}
implies that the relative dynamics in the present case can not be
properly described by the SG equation. We have also considered a much
stronger coupling energy and the result shows similar behaviour except
a much larger amplitude oscillation in the total phase. The numerical
results show the possibility to study the oscillon-type excitation
in the strongly coupled BECs. 

\subsubsection{Phase Diagram}

From the previous simulations, we conclude that the long-lived oscillon
excitation can exist in both weak- and strong coupling regimes. In
general, the stability of the oscillon depends on two parameters,
$\nu$ and $u$. We notice that $u$ is not directly related to system parameters, but it can be expressed in terms of $\nu$ and $\omega_{B}$,
the frequency of the initial SG breather. Therefore, it is more practical to characterize the stability of the oscillon as a function of the coupling $\nu$, and the frequency
$\omega_{B}$ of the imprinted breather. 
We take the relative energy, $E_{r}$, which is the ratio of remnant
GP energy to the GP energy of the imprinted initial state as a measure
of the stability of the oscillon. Here the GP energy of the oscillon
is evaluated by integrating the energy density over the interval $-7z_{w}<z<7z_{w}$.
In order to avoid the transient effects of the initial phase imprinting,
the relative energy is obtained by calculating the ratio of the GP energy
at $t=15T_{0}$ to that at $t=0$, where $T_{0}=2\pi/\omega_{imprint}$
is the period of the imprinted SG breather. In the simulation, $\nu$
is varied from 0.01 to 0.6, and for each $\nu$, the imprinted frequency $\omega_{imprint}$
is varied from 0.005 to $\sqrt{4\nu\left(1+\nu\right)}$.
The limiting value $\sqrt{4\nu\left(1+\nu\right)}$ is the maximally possible frequency of the SG breather of Eq.~(\ref{breathfreq}) and indicates the
boundary beyond which the SG breather cannot be imprinted.
Numerically we also find an energy threshold for stability given by $E_{r}\sim 0.4$, i.e., the oscillon
is unstable if $0<E_{r}\lesssim 0.4$ and stable if $0.4\lesssim E_{r}\leq1$.
\par
\begin{figure}[!t]
\includegraphics[width=1.05\columnwidth]{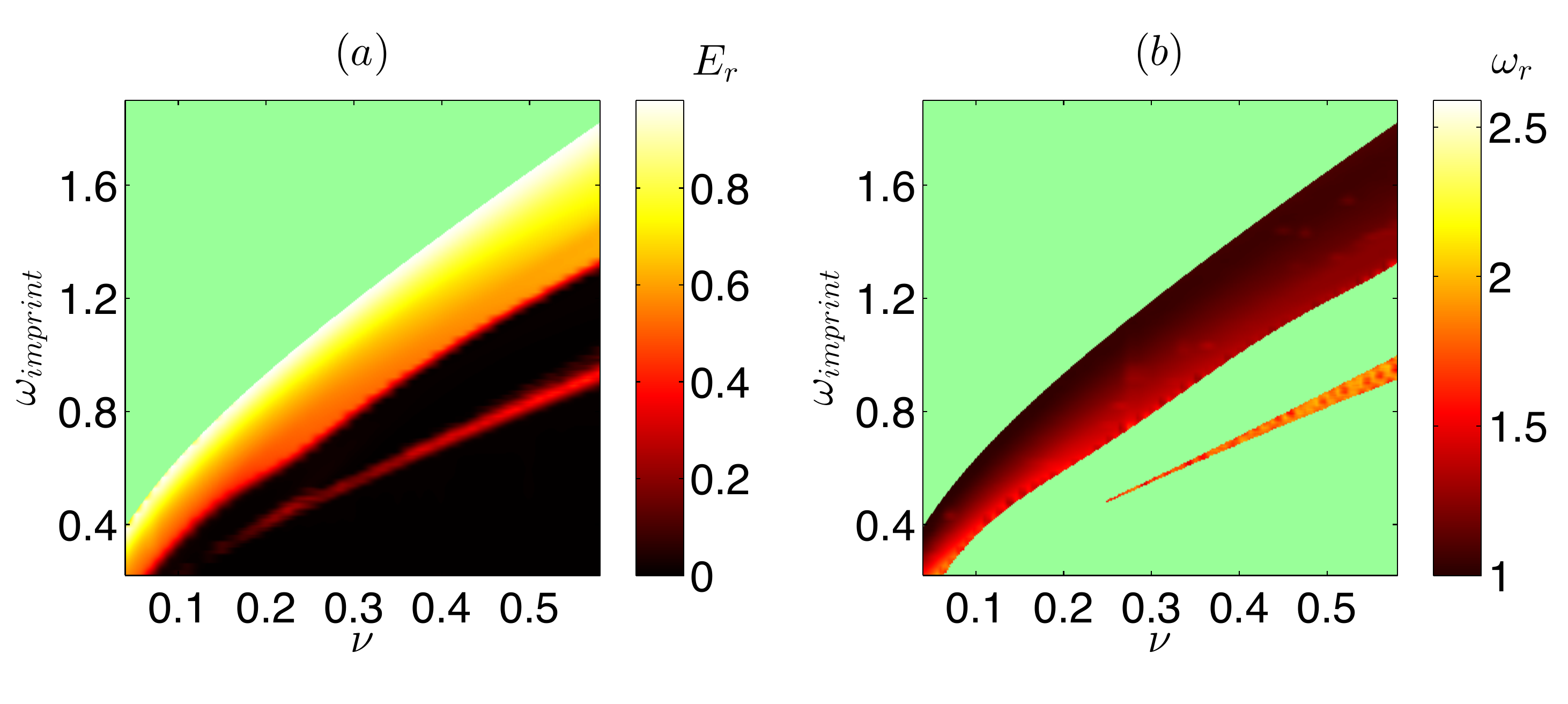}
\protect\caption{(Color online) Oscillon stability: (a) The ratio of the remnant GP
energy of the oscillon to that of the initial state at\textbf{
$t=15T_{0}$} as a function of $\omega_{imprint}$ and $\nu$ (see
text). (b) The relative frequency as a function of $\omega_{imprint}$
and $\nu$. The upper green regions in panels (a) and (b) indicate where the SG-breather initial condition cannot be imprinted. The unstable (black) region in (a) corresponds to the lower green region in (b), as $\omega_{measure}$ is undefined.}
\label{phase-diagram} 
\end{figure}

The results are shown in Fig.~\ref{phase-diagram}(a), where we see
that the oscillon as the remnant of the imprinted SG breather is always
long-lived ($E_{r}\thicksim1$) as long as $\nu<0.04$. In Fig.~\ref{phase-diagram}(a),
the upper green area marks the forbidden domain where no SG breather
can be phase-imprinted and the black area indicates the domain where
the imprinted SG breather would instantly decay into two GP solitons
as shown in Fig.~\ref{nu-5e-1-u-5e-1}. Moreover, we observe that the oscillon 
energy loss is 
accompanied by an up-chirping of its oscillation frequency with time. Hence, the frequency chirping of the oscillon can also
serve as a measure of stability. To this end, we consider the relative
frequency, $\omega_{r}=\overline{\omega}_{measure}/\omega_{imprint}$,
where $\overline{\omega}_{measure}$ is defined as the mean frequency
of the oscillon averaged over the time interval $10T_{0}\leq t\leq15T_{0}$; this interval includes several oscillations, while being much shorter than the chirp timescale.
The results of the frequency chirping are shown in Fig.~\ref{phase-diagram}(b).
Besides the green area in the energy diagram, there are two stable domains
having relative energy above the threshold. The upper stable domain
in the energy diagram shows a lower chirping rate frequency comparing
to that of the lower (narrow) stable domain. This suggests that the
energy decay is stronger in the lower stable domain than that in the
upper stable domain. We also observe that, while increasing the frequency
of the imprinted SG breather at a given coupling energy, the two grey GP
solitons, originating from the decay of the imprinted SG breather,
would gradually speed up in the black area of the energy diagram.
The decay into two grey GP solitons arises in the lower stability
region as well, yet the emitted solitons acquire a speed very close
to that of Bogoliubov sound, indicating that the solitons are very
unstable in this domain. Note that in the upper stable domain, there
are no grey solitons, but instead two outgoing small density bumps
appear when the oscillon is stablized as shown in Fig.~\ref{nu5e-1-u10}. 

We note that, since the dynamical properties of the oscillon in strong coupling regime are different from those of the SG breather, the frequency of the imprinted SG-breather may not serve as the relevant parameter to describe the stability of oscillon in strong coupling regime, which leads to the appearance of the bifurcated stable domain in the strong coupling regime (starting around $\nu\approx0.2$ in Fig.~\ref{phase-diagram}).

We have shown how the oscillon originating from a SG breather loses its energy in both weak- and strong-coupling cases. The influence of external perturbations upon the dynamics of topology-protected excitations \cite{Rooney2010}, so to speak, is broadly analogous to the influence of integrability breaking on oscillons and solitons~\cite{Martin2010,Parker2010}, namely, that the breakdown of strict integrability couples the excitation to additional degrees of freedom that can extract its energy. Intriguingly, we find that the formation of oscillons could be related to the collision of Josesphson vortices and in next section we shall address such a possibility.

\section{Formation of an oscillon} 

As Josephson vortices are topologically stable
excitations of the coupled BEC system, they can spontaneously form
during the BEC phase transition via the Kibble-Zurek mechanism~\cite{Su2013,Das},
and may then collide to form an oscillon. A Josephson vortex pair could also be
phase imprinted and then allowed to collide. In this section we study the Josephson vortex collision process as a basic prototype of oscillon formation.

\subsection{Collision of Josephson Vortices} \label{S:collison}
We simulate the collision of two Josephson vortices of opposite handedness, namely,
kink and antikink, which provides a possibility of generating an
oscillon for Eq.~(\ref{eq:DGPE-1}). 
\begin{figure}[!t]
\includegraphics[width=1\columnwidth]{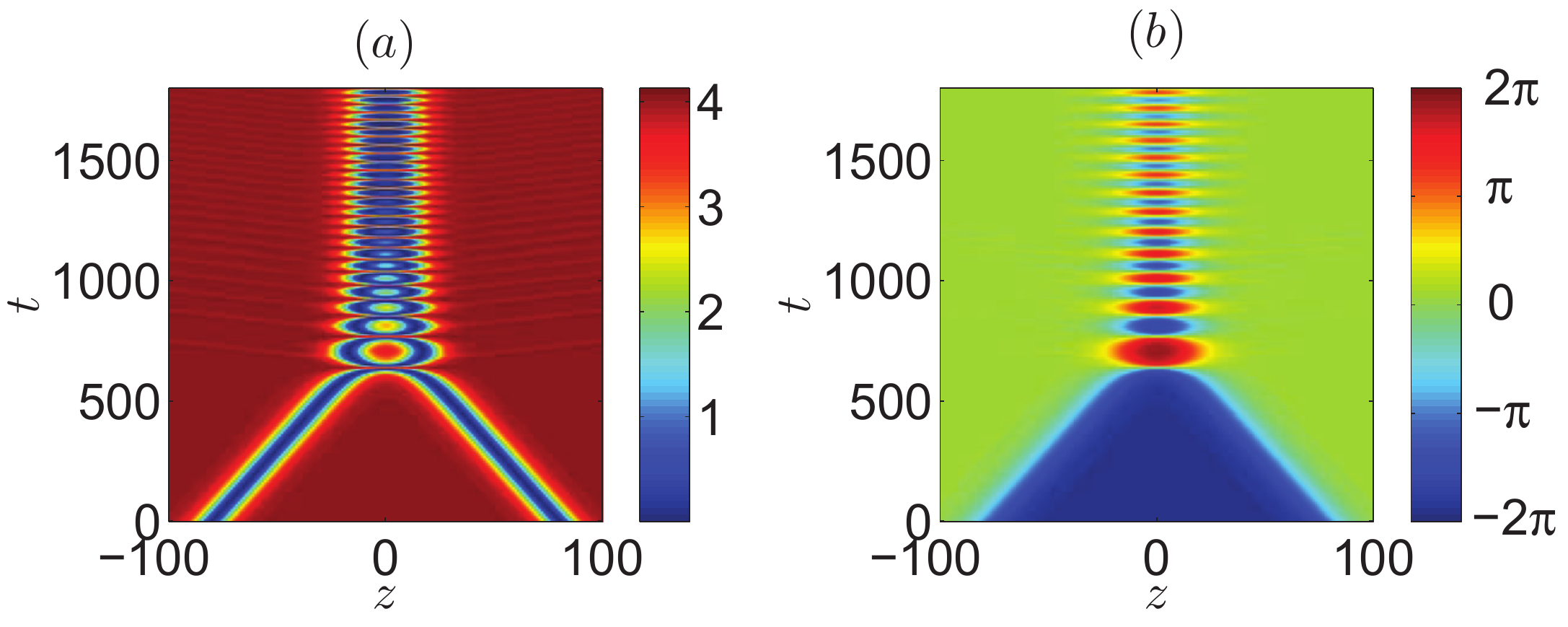}
\protect\caption{(Color online) Collision of two Josephson vortices ($\nu=0.005$): (a) The density
profile of the two superposed atomic fields, $|\psi_{1}+\psi_{2}|^{2}$.
(b) The spatial distribution of the relative phase, $\phi_{1}-\phi_{2}$.
The two counter-moving Josephson vortices with velocities $v=\pm0.1$ collide at
the instant $t=800$ forming an oscillon. As shown in panel (a), the
quasi-breather continuously radiates Bogoliubov sound waves to the
background superfluid, causing an up-chirp of the oscillon's oscillation
frequency. }
\label{collision} 
\end{figure}
We recall that the static Josephson vortices in the linearly coupled BECs exist when $\tilde{\nu}/\mu<1/3$, and closed form expressions are known from Ref.~\cite{Kaurov2005}.
To obtain a solution for moving Josephson vortices, we assume solutions propagating with constant velocity and  solve the resulting coupled GPEs numerically, increasing the velocity in small steps starting from zero~\cite{shamailov14}. The initial state for the time-dependent simulation is made up of two counter-moving Josephson vortices with speed $\left|v\right|=0.1$, initially located at both sides with equal distance from the origin.
As can be seen from
Fig.~\ref{collision}, the head-on collision of two Josephson vortices takes place
at the origin. Remarkably, the two Josephson vortices do not annihilate each other
but bind together to form a oscillon excitation after the collision.
As soon as the oscillon is formed, it continuously emits Bogoliubov
sound waves. Since the coupling is very weak, it is expected that
the relative phase dynamics is essentially governed by the SG equation,
and the oscillon emerging from the collision of two Josephson vortices (SG kinks)
appears very close to a SG breather, as the one depicted in Fig.~\ref{nu1e-2-u-5e-1}. As shown in Fig.~\ref{collision}(a), the frequency
of the oscillon increases with time. According to Eq. (\ref{Breather-energy}),
increasing $\omega_{B}$ will lower the value of $E_{B}$, implying
that the oscillon in Fig.~\ref{collision}(a) gradually loses its
energy. This decay is due to the fact that the relative phase $\phi_{a}$
can couple to the other degrees of freedom in the GP energy, namely,
the total phase $\phi_{S}$ and density $\rho$. Although the oscillon
is not an exact SG breather and undergoes damping, it has a very long
lifetime comparing to its breathing period, suggesting that oscillon
excitations in linearly coupled BECs may be observable in BEC experiments.
\begin{figure}[!t]
\includegraphics[width=1\columnwidth]{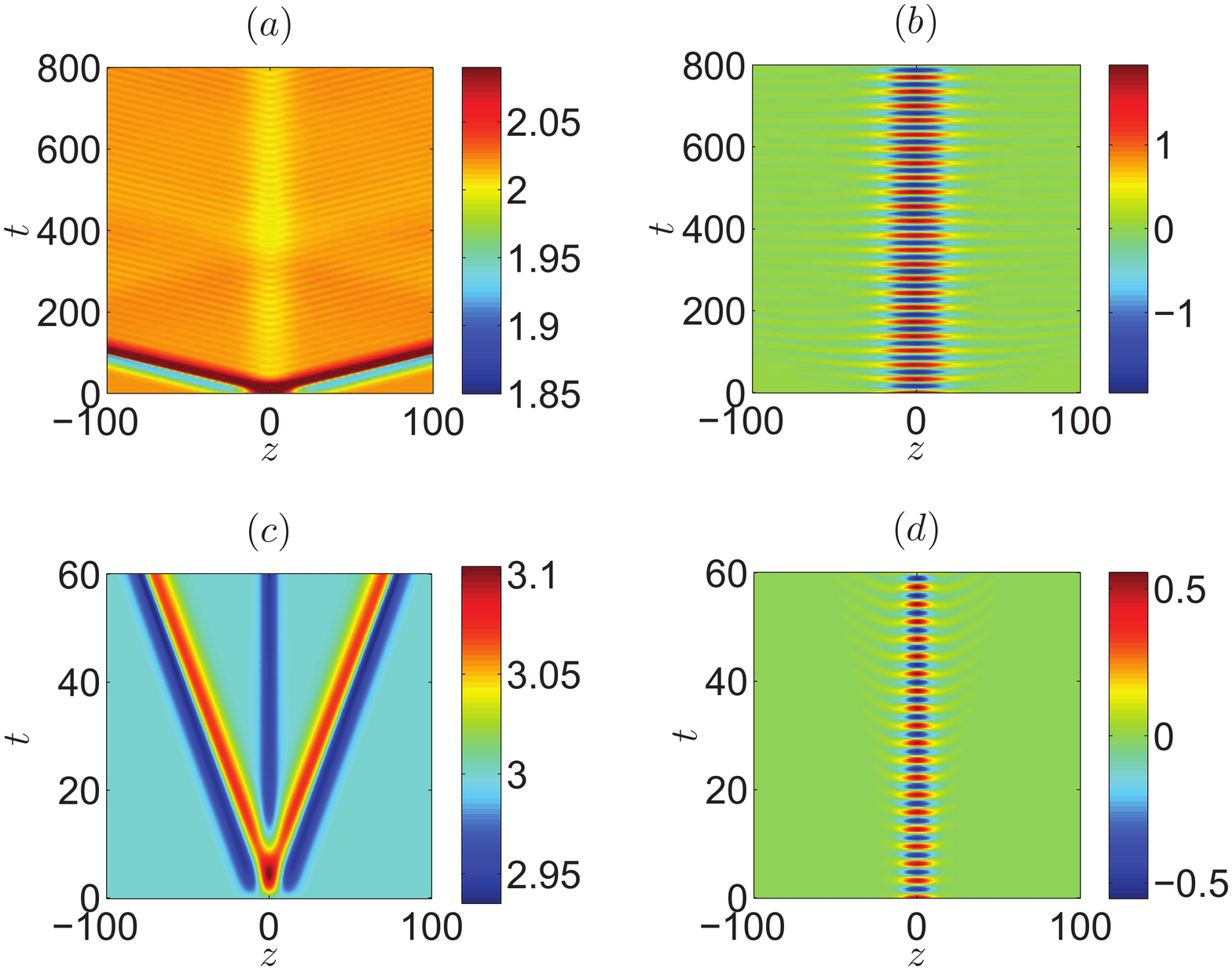}
\protect\caption{ (Color online) Dynamics following a Gaussian phase imprint as described
in the text. Panels (a) and (c) show the density profile of the superposition $|\psi_{1}+\psi_{2}|^{2}$, and (b) and (d) show the relative phase
$\phi_{1}-\phi_{2}$ for two different sets of initial parameters:
$u=2$, $\nu=0.01$, $\epsilon=0.2$ {[}(a) and (b){]} and $u=10$,
$\nu=0.5$, $\epsilon=0.2$ {[}(c) and (d){]}. In both cases a long-lived
breather is created as seen in the relative phase.}
\label{nu-1e-2-u-2-Gaussian} 
\end{figure}

\subsection{Constraints for Phase Imprinting}

We have systematically studied the oscillon in the coupled BECs at
different regimes of coupling strength. A possible way to realize
the oscillon is to phase-imprint the relative phase profile by shining
a Gaussian laser beam onto the coupled BEC system, so that the Stark
effect generated by the laser can cause a controllable position-dependent
phase difference between the two BECs. In order to create the required relative
phase profile, the laser beam should be focused between the two BECs
and the focusing point should sit slightly away from the centerline between
the two BECs. Assuming that the phases imprinted on the two BECs take
the form 
\begin{equation}
\phi_{1}^{i}=A\exp\left[-z^{2}/\kappa^{2}\right],\phi_{2}^{i}=\epsilon A\exp\left[-z^{2}/\kappa^{2}\right],\label{imprinted-phase}
\end{equation}
where $\kappa$ is the width of the Gaussian beam, $\alpha$ the geometric
factor due to the departure of focal point from the centerline between
the two condensates, and $A$ is proportional to the product of laser
intensity and time duration of the exposure. As a consequence, the
relative phase is given by $\Delta\phi=(1-\epsilon)A\exp\left[-z^{2}/\kappa^{2}\right]$.
Matching the amplitude and width of $\triangle\phi$ to those of the
SG breather given by Eq.~(\ref{breather}), the width $\kappa$ and
the amplitude $A$ can be determined after some algebra as 
\begin{eqnarray}
\kappa^{2} & = & \frac{z_{w}^{2}}{\ln(\tan^{-1}\frac{\sqrt{1+\nu}}{u})-\ln(\tan\frac{\sqrt{1+\nu}}{2u})},\label{width}\\
A & = & \frac{4}{1-\epsilon}\tan^{-1}\frac{\sqrt{1+\nu}}{u},\label{intensity}
\end{eqnarray}
where $z_{w}$ is the FWHM of the SG breather defined previously.
Note that parameter range of the imprinted Gaussian phase is limited
since Eq.~(\ref{width}) becomes imaginary when $\tan(\sqrt{1+\nu}/2u)>\tan^{-1}(\sqrt{1+\nu}/u)$.

To demonstrate that the aforementioned method is practical, we simulate
Eq.~(\ref{eq:DGPE-1}) by phase imprinting the Gaussian phase profile
with $(\nu,\epsilon,u)=(0.01,0.2,2)$ and $(0.5,0.2,10)$ and the
width $\kappa$ and the amplitude $A$ are given according to Eqs.~(\ref{width}) and (\ref{intensity}). Note that according to the
assumption in Eq.~(\ref{imprinted-phase}), a non-zero total phase
will also be imprinted, which causes disturbance in the total density.
As shown in Fig.~\ref{nu-1e-2-u-2-Gaussian}, this disturbance is
detectable only in the atomic interference pattern {[}panels (a) and
(c){]} despite the creation of an oscillon excitation in the relative
phase {[}panels (b) and (d){]}. 

\section{CONCLUSIONS}

We have theoretically studied the dynamics of an
oscillon in the relative phase of two linearly coupled Bose-Einstein
condensates. The stability of the oscillon has been examined over
broad ranges of coupling energy and imprinted SG breather frequency.
In the weak coupling limit, the system exhibits SG dynamics predominantly,
and thus an oscillon forms a long-lived metastable excitation, despite
gradually losing energy to Bogoliubov phonons. In the strong coupling
regime, the system dynamics is no longer governed by the SG equation
for the relative phase due to the increasingly prominent interplay
between the symmetric and asymmetric degrees of freedom. A systematic
study of parameter space reveals large regions where
the lifetime of the oscillon exceeds its period by several orders of magnitude. 
The experimental realisation of oscillons may enable
the possibility to use Bose-Einstein condensates as an analogue model
to simulate and characterize oscillon formation in the dynamics of
the early universe.
 
\acknowledgements
We are indebted to Sophie Shamailov for the provision of numerical code for generating moving Josephson vortices used for Sec.~\ref{S:collison} and for useful comments on the manuscript. 
AB was supported by a Rutherford Discovery Fellowship administered by the Royal Society of New Zealand. 
OF was supported by the Marsden Fund (Project No.\ MAU 1205), administrated by the Royal 
Society of New Zealand. SWS, SCG, and IKL were supported by the Ministry of Science and Technology, 
Taiwan (Grant No. MOST 103-2112- M-018-002-MY3). SCG was also supported by National Centre for Theoretical Scientists, Taiwan.


\begin{thebibliography}{0}%
\makeatletter
\providecommand \@ifxundefined [1]{%
 \@ifx{#1\undefined}
}%
\providecommand \@ifnum [1]{%
 \ifnum #1\expandafter \@firstoftwo
 \else \expandafter \@secondoftwo
 \fi
}%
\providecommand \@ifx [1]{%
 \ifx #1\expandafter \@firstoftwo
 \else \expandafter \@secondoftwo
 \fi
}%
\providecommand \natexlab [1]{#1}%
\providecommand \enquote  [1]{``#1''}%
\providecommand \bibnamefont  [1]{#1}%
\providecommand \bibfnamefont [1]{#1}%
\providecommand \citenamefont [1]{#1}%
\providecommand \href@noop [0]{\@secondoftwo}%
\providecommand \href [0]{\begingroup \@sanitize@url \@href}%
\providecommand \@href[1]{\@@startlink{#1}\@@href}%
\providecommand \@@href[1]{\endgroup#1\@@endlink}%
\providecommand \@sanitize@url [0]{\catcode `\\12\catcode `\$12\catcode
  `\&12\catcode `\#12\catcode `\^12\catcode `\_12\catcode `\%12\relax}%
\providecommand \@@startlink[1]{}%
\providecommand \@@endlink[0]{}%
\providecommand \url  [0]{\begingroup\@sanitize@url \@url }%
\providecommand \@url [1]{\endgroup\@href {#1}{\urlprefix }}%
\providecommand \urlprefix  [0]{URL }%
\providecommand \Eprint [0]{\href }%
\providecommand \doibase [0]{http://dx.doi.org/}%
\providecommand \selectlanguage [0]{\@gobble}%
\providecommand \bibinfo  [0]{\@secondoftwo}%
\providecommand \bibfield  [0]{\@secondoftwo}%
\providecommand \translation [1]{[#1]}%
\providecommand \BibitemOpen [0]{}%
\providecommand \bibitemStop [0]{}%
\providecommand \bibitemNoStop [0]{.\EOS\space}%
\providecommand \EOS [0]{\spacefactor3000\relax}%
\providecommand \BibitemShut  [1]{\csname bibitem#1\endcsname}%
\let\auto@bib@innerbib\@empty
\bibitem [{\citenamefont {Bogolyubskii}\ and\ \citenamefont
  {Makhankov}(1976)}]{Bogolyubskii1976}%
  \BibitemOpen
  \bibfield  {author} {\bibinfo {author} {\bibfnamefont {I.}~\bibnamefont
  {Bogolyubskii}}\ and\ \bibinfo {author} {\bibfnamefont {V.}~\bibnamefont
  {Makhankov}},\ }\href {http://jetpletters.ac.ru/ps/1809/article\_27642.pdf}
  {\bibfield  {journal} {\bibinfo  {journal} {JETP Lett}\ }\textbf {\bibinfo
  {volume} {24}},\ \bibinfo {pages} {12} (\bibinfo {year} {1976})}\BibitemShut
  {NoStop}%
\bibitem [{\citenamefont {Gleiser}(1994)}]{Gleiser1994}%
  \BibitemOpen
  \bibfield  {author} {\bibinfo {author} {\bibfnamefont {M.}~\bibnamefont
  {Gleiser}},\ }\href {\doibase 10.1103/PhysRevD.49.2978} {\bibfield  {journal}
  {\bibinfo  {journal} {Phys. Rev. D}\ }\textbf {\bibinfo {volume} {49}},\
  \bibinfo {pages} {2978} (\bibinfo {year} {1994})}\BibitemShut {NoStop}%
\bibitem [{\citenamefont {Vachaspati}(2006)}]{Vachaspati2006}%
  \BibitemOpen
  \bibfield  {author} {\bibinfo {author} {\bibfnamefont {T.}~\bibnamefont
  {Vachaspati}},\ }\href
  {http://books.google.com/books?id=dCkiEFVH2KMC\&pgis=1} {\emph {\bibinfo
  {title} {{Kinks and Domain Walls: An Introduction to Classical and Quantum
  Solitons}}}}\ (\bibinfo  {publisher} {Cambridge University Press},\ \bibinfo
  {address} {{UK}},\ \bibinfo {year} {2006})\BibitemShut {NoStop}%
\bibitem [{\citenamefont {Copeland}\ \emph {et~al.}(1995)\citenamefont
  {Copeland}, \citenamefont {Gleiser},\ and\ \citenamefont
  {M\"{u}ller}}]{Copeland1995}%
  \BibitemOpen
  \bibfield  {author} {\bibinfo {author} {\bibfnamefont {E.}~\bibnamefont
  {Copeland}}, \bibinfo {author} {\bibfnamefont {M.}~\bibnamefont {Gleiser}}, \
  and\ \bibinfo {author} {\bibfnamefont {H.-R.}\ \bibnamefont {M\"{u}ller}},\
  }\href {\doibase 10.1103/PhysRevD.52.1920} {\bibfield  {journal} {\bibinfo
  {journal} {Phys. Rev. D}\ }\textbf {\bibinfo {volume} {52}},\ \bibinfo
  {pages} {1920} (\bibinfo {year} {1995})}\BibitemShut {NoStop}%
\bibitem [{\citenamefont {Amin}\ \emph {et~al.}(2012)\citenamefont {Amin},
  \citenamefont {Easther}, \citenamefont {Finkel}, \citenamefont {Flauger},\
  and\ \citenamefont {Hertzberg}}]{Amin2012}%
  \BibitemOpen
  \bibfield  {author} {\bibinfo {author} {\bibfnamefont {M.~A.}\ \bibnamefont
  {Amin}}, \bibinfo {author} {\bibfnamefont {R.}~\bibnamefont {Easther}},
  \bibinfo {author} {\bibfnamefont {H.}~\bibnamefont {Finkel}}, \bibinfo
  {author} {\bibfnamefont {R.}~\bibnamefont {Flauger}}, \ and\ \bibinfo
  {author} {\bibfnamefont {M.~P.}\ \bibnamefont {Hertzberg}},\ }\href {\doibase
  10.1103/PhysRevLett.108.241302} {\bibfield  {journal} {\bibinfo  {journal}
  {Phys. Rev. Lett.}\ }\textbf {\bibinfo {volume} {108}},\ \bibinfo {pages}
  {241302} (\bibinfo {year} {2012})}\BibitemShut {NoStop}%
\bibitem [{\citenamefont {Hindmarsh}\ and\ \citenamefont
  {Kibble}(1995)}]{Hindmarsh1995}%
  \BibitemOpen
  \bibfield  {author} {\bibinfo {author} {\bibfnamefont {M.~B.}\ \bibnamefont
  {Hindmarsh}}\ and\ \bibinfo {author} {\bibfnamefont {T.~W.~B.}\ \bibnamefont
  {Kibble}},\ }\href {\doibase 10.1088/0034-4885/58/5/001} {\bibfield
  {journal} {\bibinfo  {journal} {Reports Prog. Phys.}\ }\textbf {\bibinfo
  {volume} {58}},\ \bibinfo {pages} {477} (\bibinfo {year} {1995})},\ \Eprint
  {http://arxiv.org/abs/9411342} {arXiv:9411342 [hep-ph]} \BibitemShut
  {NoStop}%
\bibitem [{\citenamefont {Ablowitz}\ \emph {et~al.}(1973)\citenamefont
  {Ablowitz}, \citenamefont {Kaup}, \citenamefont {Newell},\ and\ \citenamefont
  {Segur}}]{Ablowitz1973}%
  \BibitemOpen
  \bibfield  {author} {\bibinfo {author} {\bibfnamefont {M.}~\bibnamefont
  {Ablowitz}}, \bibinfo {author} {\bibfnamefont {D.}~\bibnamefont {Kaup}},
  \bibinfo {author} {\bibfnamefont {A.}~\bibnamefont {Newell}}, \ and\ \bibinfo
  {author} {\bibfnamefont {H.}~\bibnamefont {Segur}},\ }\href {\doibase
  10.1103/PhysRevLett.30.1262} {\bibfield  {journal} {\bibinfo  {journal}
  {Phys. Rev. Lett.}\ }\textbf {\bibinfo {volume} {30}},\ \bibinfo {pages}
  {1262} (\bibinfo {year} {1973})}\BibitemShut {NoStop}%
\bibitem [{\citenamefont {Akhmediev}\ \emph {et~al.}(1987)\citenamefont
  {Akhmediev}, \citenamefont {Eleonskii},\ and\ \citenamefont
  {Kulagin}}]{Akhmediev1987}%
  \BibitemOpen
  \bibfield  {author} {\bibinfo {author} {\bibfnamefont {N.~N.}\ \bibnamefont
  {Akhmediev}}, \bibinfo {author} {\bibfnamefont {V.~M.}\ \bibnamefont
  {Eleonskii}}, \ and\ \bibinfo {author} {\bibfnamefont {N.~E.}\ \bibnamefont
  {Kulagin}},\ }\href {\doibase 10.1007/BF01017105} {\bibfield  {journal}
  {\bibinfo  {journal} {Theor. Math. Phys.}\ }\textbf {\bibinfo {volume}
  {72}},\ \bibinfo {pages} {809} (\bibinfo {year} {1987})}\BibitemShut
  {NoStop}%
\bibitem [{\citenamefont {Flach}\ and\ \citenamefont
  {Willis}(1998)}]{Flach1998}%
  \BibitemOpen
  \bibfield  {author} {\bibinfo {author} {\bibfnamefont {S.}~\bibnamefont
  {Flach}}\ and\ \bibinfo {author} {\bibfnamefont {C.}~\bibnamefont {Willis}},\
  }\href {\doibase 10.1016/S0370-1573(97)00068-9} {\bibfield  {journal}
  {\bibinfo  {journal} {Phys. Rep.}\ }\textbf {\bibinfo {volume} {295}},\
  \bibinfo {pages} {181} (\bibinfo {year} {1998})}\BibitemShut {NoStop}%
\bibitem [{\citenamefont {Bour}(1862)}]{Bour}%
  \BibitemOpen
  \bibfield  {author} {\bibinfo {author} {\bibfnamefont {E.}~\bibnamefont
  {Bour}},\ }\href@noop {} {\bibfield  {journal} {\bibinfo  {journal} {Journal
  de l'\'{e}cole Imperiale Polytechnique}\ } (\bibinfo {year}
  {1862})}\BibitemShut {NoStop}%
\bibitem [{\citenamefont {Frenkel}\ and\ \citenamefont
  {Kontrova}(1939)}]{Frenkel}%
  \BibitemOpen
  \bibfield  {author} {\bibinfo {author} {\bibfnamefont {J.}~\bibnamefont
  {Frenkel}}\ and\ \bibinfo {author} {\bibfnamefont {T.}~\bibnamefont
  {Kontrova}},\ }\href@noop {} {\bibfield  {journal} {\bibinfo  {journal} {J.
  Phys.}\ }\textbf {\bibinfo {volume} {1}},\ \bibinfo {pages} {137} (\bibinfo
  {year} {1939})}\BibitemShut {NoStop}%
\bibitem [{\citenamefont {Dauxois}\ and\ \citenamefont
  {Peyrard}(2006)}]{Physofsoliton}%
  \BibitemOpen
  \bibfield  {author} {\bibinfo {author} {\bibfnamefont {T.}~\bibnamefont
  {Dauxois}}\ and\ \bibinfo {author} {\bibfnamefont {M.}~\bibnamefont
  {Peyrard}},\ }\href@noop {} {\emph {\bibinfo {title} {{Physics of
  Solitons}}}}\ (\bibinfo  {publisher} {{Cambridge University Press}},\
  \bibinfo {address} {{UK}},\ \bibinfo {year} {2006})\BibitemShut {NoStop}%
\bibitem [{\citenamefont {Peyrard}\ and\ \citenamefont
  {Campbell}(1983)}]{Peyrard1983}%
  \BibitemOpen
  \bibfield  {author} {\bibinfo {author} {\bibfnamefont {M.}~\bibnamefont
  {Peyrard}}\ and\ \bibinfo {author} {\bibfnamefont {D.~K.}\ \bibnamefont
  {Campbell}},\ }\href {\doibase 10.1016/0167-2789(83)90290-7} {\bibfield
  {journal} {\bibinfo  {journal} {Phys. D Nonlinear Phenom.}\ }\textbf
  {\bibinfo {volume} {9}},\ \bibinfo {pages} {33} (\bibinfo {year}
  {1983})}\BibitemShut {NoStop}%
\bibitem [{\citenamefont {Abdullaev}\ \emph {et~al.}(2013)\citenamefont
  {Abdullaev}, \citenamefont {Gammal}, \citenamefont {Malomed},\ and\
  \citenamefont {Tomio}}]{Abdullaev2013}%
  \BibitemOpen
  \bibfield  {author} {\bibinfo {author} {\bibfnamefont {F.~K.}\ \bibnamefont
  {Abdullaev}}, \bibinfo {author} {\bibfnamefont {a.}~\bibnamefont {Gammal}},
  \bibinfo {author} {\bibfnamefont {B.~a.}\ \bibnamefont {Malomed}}, \ and\
  \bibinfo {author} {\bibfnamefont {L.}~\bibnamefont {Tomio}},\ }\href
  {\doibase 10.1103/PhysRevA.87.063621} {\bibfield  {journal} {\bibinfo
  {journal} {Physical Review A}\ }\textbf {\bibinfo {volume} {87}},\ \bibinfo
  {pages} {063621} (\bibinfo {year} {2013})}\BibitemShut {NoStop}%
\bibitem [{\citenamefont {Gritsev}\ \emph {et~al.}(2007)\citenamefont
  {Gritsev}, \citenamefont {Polkovnikov},\ and\ \citenamefont
  {Demler}}]{Gritsev2007}%
  \BibitemOpen
  \bibfield  {author} {\bibinfo {author} {\bibfnamefont {V.}~\bibnamefont
  {Gritsev}}, \bibinfo {author} {\bibfnamefont {A.}~\bibnamefont
  {Polkovnikov}}, \ and\ \bibinfo {author} {\bibfnamefont {E.}~\bibnamefont
  {Demler}},\ }\href {\doibase 10.1103/PhysRevB.75.174511} {\bibfield
  {journal} {\bibinfo  {journal} {Phys. Rev. B}\ }\textbf {\bibinfo {volume}
  {75}},\ \bibinfo {pages} {174511} (\bibinfo {year} {2007})}\BibitemShut
  {NoStop}%
\bibitem [{\citenamefont {Neuenhahn}\ \emph {et~al.}(2012)\citenamefont
  {Neuenhahn}, \citenamefont {Polkovnikov},\ and\ \citenamefont
  {Marquardt}}]{Neuenhahn2012}%
  \BibitemOpen
  \bibfield  {author} {\bibinfo {author} {\bibfnamefont {C.}~\bibnamefont
  {Neuenhahn}}, \bibinfo {author} {\bibfnamefont {A.}~\bibnamefont
  {Polkovnikov}}, \ and\ \bibinfo {author} {\bibfnamefont {F.}~\bibnamefont
  {Marquardt}},\ }\href {\doibase 10.1103/PhysRevLett.109.085304} {\bibfield
  {journal} {\bibinfo  {journal} {Phys. Rev. Lett.}\ }\textbf {\bibinfo
  {volume} {109}},\ \bibinfo {pages} {085304} (\bibinfo {year}
  {2012})}\BibitemShut {NoStop}%
\bibitem [{\citenamefont {Opanchuk}\ \emph {et~al.}(2013)\citenamefont
  {Opanchuk}, \citenamefont {Polkinghorne}, \citenamefont {Fialko},
  \citenamefont {Brand},\ and\ \citenamefont {Drummond}}]{Opanchuk2013}%
  \BibitemOpen
  \bibfield  {author} {\bibinfo {author} {\bibfnamefont {B.}~\bibnamefont
  {Opanchuk}}, \bibinfo {author} {\bibfnamefont {R.}~\bibnamefont
  {Polkinghorne}}, \bibinfo {author} {\bibfnamefont {O.}~\bibnamefont
  {Fialko}}, \bibinfo {author} {\bibfnamefont {J.}~\bibnamefont {Brand}}, \
  and\ \bibinfo {author} {\bibfnamefont {P.~D.}\ \bibnamefont {Drummond}},\
  }\href {\doibase 10.1002/andp.201300113} {\bibfield  {journal} {\bibinfo
  {journal} {Ann. Phys.}\ }\textbf {\bibinfo {volume} {525}},\ \bibinfo {pages}
  {866} (\bibinfo {year} {2013})},\ \Eprint {http://arxiv.org/abs/1305.5314}
  {arXiv:1305.5314} \BibitemShut {NoStop}%
\bibitem [{\citenamefont {Kaurov}\ and\ \citenamefont
  {Kuklov}(2005)}]{Kaurov2005}%
  \BibitemOpen
  \bibfield  {author} {\bibinfo {author} {\bibfnamefont {V.}~\bibnamefont
  {Kaurov}}\ and\ \bibinfo {author} {\bibfnamefont {A.}~\bibnamefont
  {Kuklov}},\ }\href {\doibase 10.1103/PhysRevA.71.011601} {\bibfield
  {journal} {\bibinfo  {journal} {Phys. Rev. A}\ }\textbf {\bibinfo {volume}
  {71}},\ \bibinfo {pages} {011601} (\bibinfo {year} {2005})}\BibitemShut
  {NoStop}%
\bibitem [{\citenamefont {Kaurov}\ and\ \citenamefont
  {Kuklov}(2006)}]{Kaurov2006}%
  \BibitemOpen
  \bibfield  {author} {\bibinfo {author} {\bibfnamefont {V.}~\bibnamefont
  {Kaurov}}\ and\ \bibinfo {author} {\bibfnamefont {A.}~\bibnamefont
  {Kuklov}},\ }\href {\doibase 10.1103/PhysRevA.73.013627} {\bibfield
  {journal} {\bibinfo  {journal} {Phys. Rev. A}\ }\textbf {\bibinfo {volume}
  {73}},\ \bibinfo {pages} {013627} (\bibinfo {year} {2006})}\BibitemShut
  {NoStop}%
\bibitem [{\citenamefont {Brand}\ \emph {et~al.}(2009)\citenamefont {Brand},
  \citenamefont {Haigh},\ and\ \citenamefont {Z\"{u}licke}}]{Brand2009}%
  \BibitemOpen
  \bibfield  {author} {\bibinfo {author} {\bibfnamefont {J.}~\bibnamefont
  {Brand}}, \bibinfo {author} {\bibfnamefont {T.~J.}\ \bibnamefont {Haigh}}, \
  and\ \bibinfo {author} {\bibfnamefont {U.}~\bibnamefont {Z\"{u}licke}},\
  }\href {\doibase 10.1103/PhysRevA.80.011602} {\bibfield  {journal} {\bibinfo
  {journal} {Phys. Rev. A}\ }\textbf {\bibinfo {volume} {80}},\ \bibinfo
  {pages} {011602(R)} (\bibinfo {year} {2009})}\BibitemShut {NoStop}%
\bibitem [{\citenamefont {Qadir}\ \emph {et~al.}(2012)\citenamefont {Qadir},
  \citenamefont {Susanto},\ and\ \citenamefont {Matthews}}]{Qadir2012}%
  \BibitemOpen
  \bibfield  {author} {\bibinfo {author} {\bibfnamefont {M.~I.}\ \bibnamefont
  {Qadir}}, \bibinfo {author} {\bibfnamefont {H.}~\bibnamefont {Susanto}}, \
  and\ \bibinfo {author} {\bibfnamefont {P.~C.}\ \bibnamefont {Matthews}},\
  }\href {\doibase 10.1088/0953-4075/45/3/035004} {\bibfield  {journal}
  {\bibinfo  {journal} {J. Phys. B At. Mol. Opt. Phys.}\ }\textbf {\bibinfo
  {volume} {45}},\ \bibinfo {pages} {035004} (\bibinfo {year}
  {2012})}\BibitemShut {NoStop}%
\bibitem [{\citenamefont {Shamailov}\ and\ \citenamefont
  {Brand}(2015)}]{shamailov14}%
  \BibitemOpen
  \bibfield  {author} {\bibinfo {author} {\bibfnamefont {S.~S.}\ \bibnamefont
  {Shamailov}}\ and\ \bibinfo {author} {\bibfnamefont {J.}~\bibnamefont
  {Brand}},\ }\href@noop {} {}\bibinfo {howpublished} {to be published}
  (\bibinfo {year} {2015})\BibitemShut {NoStop}%
\bibitem [{\citenamefont {Su}\ \emph {et~al.}(2013)\citenamefont {Su},
  \citenamefont {Gou}, \citenamefont {Bradley}, \citenamefont {Fialko},\ and\
  \citenamefont {Brand}}]{Su2013}%
  \BibitemOpen
  \bibfield  {author} {\bibinfo {author} {\bibfnamefont {S.-W.}\ \bibnamefont
  {Su}}, \bibinfo {author} {\bibfnamefont {S.-C.}\ \bibnamefont {Gou}},
  \bibinfo {author} {\bibfnamefont {A.}~\bibnamefont {Bradley}}, \bibinfo
  {author} {\bibfnamefont {O.}~\bibnamefont {Fialko}}, \ and\ \bibinfo {author}
  {\bibfnamefont {J.}~\bibnamefont {Brand}},\ }\href {\doibase
  10.1103/PhysRevLett.110.215302} {\bibfield  {journal} {\bibinfo  {journal}
  {Phys. Rev. Lett.}\ }\textbf {\bibinfo {volume} {110}},\ \bibinfo {pages}
  {215302} (\bibinfo {year} {2013})},\ \Eprint {http://arxiv.org/abs/1302.3304}
  {arXiv:1302.3304} \BibitemShut {NoStop}%
\bibitem [{\citenamefont {Ambegaokar}\ \emph {et~al.}(1982)\citenamefont
  {Ambegaokar}, \citenamefont {Eckern},\ and\ \citenamefont
  {Sch\"on}}]{Ambegaokar}%
  \BibitemOpen
  \bibfield  {author} {\bibinfo {author} {\bibfnamefont {V.}~\bibnamefont
  {Ambegaokar}}, \bibinfo {author} {\bibfnamefont {U.}~\bibnamefont {Eckern}},
  \ and\ \bibinfo {author} {\bibfnamefont {G.}~\bibnamefont {Sch\"on}},\
  }\href@noop {} {\bibfield  {journal} {\bibinfo  {journal} {Phys. Rev. Lett.}\
  }\textbf {\bibinfo {volume} {48}},\ \bibinfo {pages} {1745} (\bibinfo {year}
  {1982})}\BibitemShut {NoStop}%
\bibitem [{\citenamefont {Eckern}\ \emph {et~al.}(1984)\citenamefont {Eckern},
  \citenamefont {Sch\"on},\ and\ \citenamefont {Ambegaokar}}]{Eckern}%
  \BibitemOpen
  \bibfield  {author} {\bibinfo {author} {\bibfnamefont {U.}~\bibnamefont
  {Eckern}}, \bibinfo {author} {\bibfnamefont {G.}~\bibnamefont {Sch\"on}}, \
  and\ \bibinfo {author} {\bibfnamefont {V.}~\bibnamefont {Ambegaokar}},\
  }\href@noop {} {\bibfield  {journal} {\bibinfo  {journal} {Phys. Rev. B}\
  }\textbf {\bibinfo {volume} {30}},\ \bibinfo {pages} {6419} (\bibinfo {year}
  {1984})}\BibitemShut {NoStop}%
\bibitem [{\citenamefont {Segur}\ and\ \citenamefont {Kruskal}(1987)}]{Harvey}%
  \BibitemOpen
  \bibfield  {author} {\bibinfo {author} {\bibfnamefont {H.}~\bibnamefont
  {Segur}}\ and\ \bibinfo {author} {\bibfnamefont {M.~D.}\ \bibnamefont
  {Kruskal}},\ }\href@noop {} {\bibfield  {journal} {\bibinfo  {journal} {Phys.
  Rev. Lett.}\ }\textbf {\bibinfo {volume} {58}},\ \bibinfo {pages} {747}
  (\bibinfo {year} {1987})}\BibitemShut {NoStop}%
\bibitem [{\citenamefont {Boyd}(1990)}]{Boyd}%
  \BibitemOpen
  \bibfield  {author} {\bibinfo {author} {\bibfnamefont {J.~P.}\ \bibnamefont
  {Boyd}},\ }\href@noop {} {\bibfield  {journal} {\bibinfo  {journal}
  {Nonlinearity}\ }\textbf {\bibinfo {volume} {3}},\ \bibinfo {pages} {177}
  (\bibinfo {year} {1990})}\BibitemShut {NoStop}%
\bibitem [{\citenamefont {Denzler}(1993)}]{Denzler}%
  \BibitemOpen
  \bibfield  {author} {\bibinfo {author} {\bibfnamefont {J.}~\bibnamefont
  {Denzler}},\ }\href@noop {} {\bibfield  {journal} {\bibinfo  {journal}
  {Communications in Mathematical physics}\ }\textbf {\bibinfo {volume}
  {158}},\ \bibinfo {pages} {397} (\bibinfo {year} {1993})}\BibitemShut
  {NoStop}%
\bibitem [{\citenamefont {Lomdahl}\ \emph {et~al.}(1984)\citenamefont
  {Lomdahl}, \citenamefont {Olsen},\ and\ \citenamefont
  {Samuelsen}}]{Samuelsen}%
  \BibitemOpen
  \bibfield  {author} {\bibinfo {author} {\bibfnamefont {P.~S.}\ \bibnamefont
  {Lomdahl}}, \bibinfo {author} {\bibfnamefont {O.~H.}\ \bibnamefont {Olsen}},
  \ and\ \bibinfo {author} {\bibfnamefont {M.~R.}\ \bibnamefont {Samuelsen}},\
  }\href@noop {} {\bibfield  {journal} {\bibinfo  {journal} {Phys. Rev. A}\
  }\textbf {\bibinfo {volume} {29}},\ \bibinfo {pages} {350} (\bibinfo {year}
  {1984})}\BibitemShut {NoStop}%
\bibitem [{\citenamefont {Penckwitt}\ \emph {et~al.}(2002)\citenamefont
  {Penckwitt}, \citenamefont {Ballabh},\ and\ \citenamefont
  {Gardiner}}]{Penckwitt}%
  \BibitemOpen
  \bibfield  {author} {\bibinfo {author} {\bibfnamefont {A.~A.}\ \bibnamefont
  {Penckwitt}}, \bibinfo {author} {\bibfnamefont {R.~J.}\ \bibnamefont
  {Ballabh}}, \ and\ \bibinfo {author} {\bibfnamefont {C.~W.}\ \bibnamefont
  {Gardiner}},\ }\href@noop {} {\bibfield  {journal} {\bibinfo  {journal}
  {Phys. Rev. Lett.}\ }\textbf {\bibinfo {volume} {89}},\ \bibinfo {pages}
  {260402} (\bibinfo {year} {2002})}\BibitemShut {NoStop}%
\bibitem [{\citenamefont {Tsubota}\ \emph {et~al.}(2002)\citenamefont
  {Tsubota}, \citenamefont {Kasamatsu},\ and\ \citenamefont {Ueda}}]{Tsubota}%
  \BibitemOpen
  \bibfield  {author} {\bibinfo {author} {\bibfnamefont {M.}~\bibnamefont
  {Tsubota}}, \bibinfo {author} {\bibfnamefont {K.}~\bibnamefont {Kasamatsu}},
  \ and\ \bibinfo {author} {\bibfnamefont {M.}~\bibnamefont {Ueda}},\
  }\href@noop {} {\bibfield  {journal} {\bibinfo  {journal} {Phys. Rev. Lett.}\
  }\textbf {\bibinfo {volume} {65}},\ \bibinfo {pages} {023603} (\bibinfo
  {year} {2002})}\BibitemShut {NoStop}%
\bibitem [{\citenamefont {Billam}\ \emph {et~al.}(2014)\citenamefont {Billam},
  \citenamefont {Reeves}, \citenamefont {Anderson},\ and\ \citenamefont
  {Bradley}}]{Billam}%
  \BibitemOpen
  \bibfield  {author} {\bibinfo {author} {\bibfnamefont {T.~P.}\ \bibnamefont
  {Billam}}, \bibinfo {author} {\bibfnamefont {M.~T.}\ \bibnamefont {Reeves}},
  \bibinfo {author} {\bibfnamefont {B.~P.}\ \bibnamefont {Anderson}}, \ and\
  \bibinfo {author} {\bibfnamefont {A.~S.}\ \bibnamefont {Bradley}},\
  }\href@noop {} {\bibfield  {journal} {\bibinfo  {journal} {Phys. Rev. Lett.}\
  }\textbf {\bibinfo {volume} {112}},\ \bibinfo {pages} {145301} (\bibinfo
  {year} {2014})}\BibitemShut {NoStop}%
\bibitem [{\citenamefont {Rooney}\ \emph {et~al.}(2010)\citenamefont {Rooney},
  \citenamefont {Bradley},\ and\ \citenamefont {Blakie}}]{Rooney2010}%
  \BibitemOpen
  \bibfield  {author} {\bibinfo {author} {\bibfnamefont {S.~J.}\ \bibnamefont
  {Rooney}}, \bibinfo {author} {\bibfnamefont {A.~S.}\ \bibnamefont {Bradley}},
  \ and\ \bibinfo {author} {\bibfnamefont {P.~B.}\ \bibnamefont {Blakie}},\
  }\href {\doibase 10.1103/PhysRevA.81.023630} {\bibfield  {journal} {\bibinfo
  {journal} {Phys. Rev. A}\ }\textbf {\bibinfo {volume} {81}},\ \bibinfo
  {pages} {023630} (\bibinfo {year} {2010})}\BibitemShut {NoStop}%
\bibitem [{\citenamefont {Martin}\ and\ \citenamefont
  {Ruostekoski}(2010)}]{Martin2010}%
  \BibitemOpen
  \bibfield  {author} {\bibinfo {author} {\bibfnamefont {A.~D.}\ \bibnamefont
  {Martin}}\ and\ \bibinfo {author} {\bibfnamefont {J.}~\bibnamefont
  {Ruostekoski}},\ }\href {\doibase 10.1088/1367-2630/12/5/055018} {\bibfield
  {journal} {\bibinfo  {journal} {New J. Phys.}\ }\textbf {\bibinfo {volume}
  {12}},\ \bibinfo {pages} {055018} (\bibinfo {year} {2010})}\BibitemShut
  {NoStop}%
\bibitem [{\citenamefont {Parker}\ \emph {et~al.}(2010)\citenamefont {Parker},
  \citenamefont {Proukakis},\ and\ \citenamefont {Adams}}]{Parker2010}%
  \BibitemOpen
  \bibfield  {author} {\bibinfo {author} {\bibfnamefont {N.~G.}\ \bibnamefont
  {Parker}}, \bibinfo {author} {\bibfnamefont {N.~P.}\ \bibnamefont
  {Proukakis}}, \ and\ \bibinfo {author} {\bibfnamefont {C.~S.}\ \bibnamefont
  {Adams}},\ }\href {\doibase 10.1103/PhysRevA.81.033606} {\bibfield  {journal}
  {\bibinfo  {journal} {Phys. Rev. A}\ }\textbf {\bibinfo {volume} {81}},\
  \bibinfo {pages} {033606} (\bibinfo {year} {2010})}\BibitemShut {NoStop}%
\bibitem [{\citenamefont {Das}\ \emph {et~al.}(2012)\citenamefont {Das},
  \citenamefont {Sabbatini},\ and\ \citenamefont {Zurek}}]{Das}%
  \BibitemOpen
  \bibfield  {author} {\bibinfo {author} {\bibfnamefont {A.}~\bibnamefont
  {Das}}, \bibinfo {author} {\bibfnamefont {J.}~\bibnamefont {Sabbatini}}, \
  and\ \bibinfo {author} {\bibfnamefont {H.~W.}\ \bibnamefont {Zurek}},\
  }\href@noop {} {\bibfield  {journal} {\bibinfo  {journal} {Scientific
  Reports}\ }\textbf {\bibinfo {volume} {2}},\ \bibinfo {pages} {352} (\bibinfo
  {year} {2012})}\BibitemShut {NoStop}%
\end{thebibliography}

%

\end{document}